\documentclass{PoS}

\usepackage{multirow}
\usepackage{amsmath}

\usepackage{lineno}
\newcommand{\juno}{\textsc{Juno}} 
\newcommand{\nue}{$\mathrm{\nu_{e}}$} 
\newcommand{\anue}{$\mathrm{\overline{\nu}_{e}}$} 
\newcommand{\pmt}{\textsc{pmt}}

\title{The Jiangmen Underground Neutrino Observatory}

\ShortTitle{The Jiangmen Underground Neutrino Observatory}

\author{\speaker{Marco GRASSI}%
         \thanks{MG acknowledges support from the Chinese Academy of Sciences President's International Fellowship Initiative grant 2015PM007.}\\
        IHEP, Chinese Academy of Sciences, Beijing, P.R.China\\
        E-mail: \email{mgrassi@ihep.ac.cn}\\
        on behalf of the \juno{} Collaboration}

\abstract{The Jiangmen Underground Neutrino Observatory (\juno{}) is a large and high precision liquid scintillator detector
under construction in the south of China. With its 20~kt target mass, it aims to achieve an unprecedented 3\% 
energy resolution at 1~MeV. Its main goal is to study the disappearance of reactor antineutrino to determine the
neutrino mass ordering, and to precisely measure the mixing parameters $\theta_{12}$, $\Delta m^2_{12}$, and
$\Delta m ^2_{ee}$. It also aims to detect neutrinos emitted from radioactive processes taking place within the inner layers
of the Earth (geoneutrinos), as well as neutrinos produced during rare supernova bursts. Neutrinos emitted in solar 
nuclear reactions could also be observed, if stringent radiopurity requirements on the scintillator are met. This manuscript
provides some highlights of \juno{}'s Physics Programme, and describes the detector design, as well as the ongoing detector R\&D.}

\FullConference{XIII International Conference on Heavy Quarks and Leptons\\
		22-27 May, 2016\\
		Blacksburg, Virginia, USA}

\begin{document}

\section{Introduction}
The Jiangmen Underground Neutrino Observatory (\juno{}) is a Liquid Scintillator Antineutrino Detector (LAND) currently under construction in the south of China (Jiangmen city, Guangdong province).
Once completed, it will be the largest LAND ever built, consisting in a 20~kt target mass made of Linear Alkyl-Benzene (LAB) liquid scintillator (LS),
monitored by roughly 17000 twenty-inch high-quantum efficiency (QE) photomultipliers (\pmt{}s) providing a \textasciitilde78\% photocoverage. 
Large photocoverage and large QE are two pivotal parameters of the experiment, which allow an unprecedented 3\% energy resolution at 1~MeV.
The conceptual design report~\cite{CDR} foresees the LS to be contained in an acrylic sphere 12~cm thick and 35.4~m wide, and the whole detector to be immersed in a cylindrical water pool, acting both as a moderator for the environmental radioactivity, and as 
a Cherenkov detector to tag and veto cosmic muons.
The ultimate control and minimal impact of calorimetry systematics is of maximal importance
to achieve the aforementioned energy resolution.
For this reason, a novel LAND design was introduced, where a 
second layer of small \pmt{}s is used to provide a second calorimetry handle with complementary systematic budget, 
allowing a combined, more precise and accurate energy scale definition. 
This calorimetry redundancy system is still under optimisation considerations in the context of \juno{} physics.

\juno{}'s main physics goal is to study neutrino oscillations by detecting reactor $\overline{\nu}_e$ produced by
two nuclear power plants, both 53~km distant from the detector, with a total nominal power of 
$36~\mathrm{GW}_{\mathrm{th}}$ (75\% of which is scheduled to be available at the beginning of the data taking). \juno{} will
be the first LAND to observe simultaneously both solar and atmospheric oscillations, and it aims to determine the 
neutrino mass ordering through their interference. \juno{} is also expected to measure several PMNS oscillation parameters~\cite{pmns}
with a precision better than 1\%. Moreover, \juno{} has a rich physics programme focused on neutrinos not originating from
reactors. For the sake of brevity, here we review neutrinos from supernova (SN) burst, solar neutrinos, and geoneutrinos, 
but a complete description of \textsc{Juno}'s physics goals can be found in~\cite{yellow_book}.

\section{Neutrino Physics at JUNO}

\noindent \textsc{ \textbf{Reactor Neutrinos}}\\
\indent Nuclear power plants are pure, powerful and isotropic sources of $\overline{\nu}_e$. Recent and past
 experiments located at different baselines with respect to nuclear power plants successfully exploited the detection of \anue{} to 
 prove neutrino oscillation, and to precisely measure some of the oscillation parameters~\cite{kamland, db, dc, reno}.
 \juno{} aims to detect \anue{} coming from two of the most powerful nuclear power plants in the world ---Taishan and Yiangjiang, located in the
 Chinese province of Guangdong--- to perform a precise measurement of their survival probability as a function of their energy.
 \anue{} detection takes place through the Inverse Beta Decay (IBD) process, where the \anue{} interacts with a proton of the liquid scintillator 
 yielding a positron and a neutron, namely 
 $$ \overline{\nu}_e + p \to e^{+} + n.$$ The positron loses all its energy via ionisation, and eventually annihilates. Such energy 
 deposition happens instantaneously, i.e. in a timescale much shorter than the typical LS deexcitation, and is called \textit{prompt}.
 On the contrary, the neutron loses its energy via thermal interactions, and eventually gets captured by an hydrogen of the LS 
 with a $\tau \sim 200~\mu$s (\textit{delayed} energy deposition), yielding a 2.2~MeV gamma. Because of the mass difference between 
 the neutron and the positron, the latter retains most of the \anue{} momentum, and the prompt energy is an accurate proxy for the 
 \anue{} energy. The delayed energy deposition is a unique signal signature, 
 and allows a powerful background rejection by means of a time and vertex correlation with the prompt energy deposition.
 
 Natural radioactivity makes up for the largest background contribution. Most of it originates from the \pmt{} glass and from the outer wall 
 of the detector, and is effectively reduced by applying a 0.5~m fiducial volume cut. Residual natural radioactivity events might occasionally fall within the 
 time and vertex windows used for signal selection, yielding \textit{accidental} coincidences which are an irreducible background. $\alpha$ radioactivity resulting
 in neutrons being ejected from stable nuclei is counted separately, and is referred to as $(\alpha,n)$. Geoneutrinos are considered 
 a background from the point of view of reactor physics, and need to be properly subtracted. The remaining backgrounds originate
 from the spallation of cosmogenic muons on LS molecules. They comprise long lived isotopes ($^9$Li and $^8$He) decaying $\beta-n$,
 and fast neutrons. Signal and background daily rates are summarised in Table ~\ref{tab:sigbkg}.
 
 The black line in Fig.~1 shows the expected prompt energy spectrum. It features two distinct oscillation patterns, a slow one arising 
 from the solar mass splitting, and a fast one arising from the atmospheric mass splitting. The neutrino mass ordering information
 is embedded in the interference between the two oscillation modes, and can be extracted using either a Fourier-based or a 
 $\chi^2$-based analysis. In the following we consider the latter (a description of the former can be found in~\cite{Zhan:2009rs}), 
 and we quantify the sensitivity to the mass ordering
 as the $\Delta \chi^2$ obtained from fitting the prompt energy spectrum under the two hypotheses of normal and inverted mass ordering:
 $$\Delta \chi^2 = |\chi^2_{min}(\mathrm{normal}) - \chi^2_{min}(\mathrm{inverted})|.$$ 
 Such a fit strongly relies on the capability to resolve the fast oscillation pattern, which in turn depends on both detector energy resolution 
 and statistics. Fig.~2 shows $\Delta \chi^2$  contours under different assumptions of these parameters. Six years of data with an energy
 resolution of 3\% at 1~MeV  allow to reach  $\Delta \chi^2 = 10$ by fitting \juno{}'s data alone. If we further constrain the fit using 
 a prospective 1.5\% determination of $\Delta_{\mu\mu}$ at \textsc{Minos}~\cite{minos}, $\Delta \chi^2$ increases up to 14.
 
\begin{table}[t]
\centering
\begin{tabular}{|c|c|c|c|c|c|c|c|}\hline\hline
Selection & IBD efficiency & IBD & Geo-$\nu$s & Accidental & $^9$Li/$^8$He & Fast $n$ & $(\alpha, n)$ \\ \hline
- & - & 83 & 1.5 & $\sim5.7\times10^4$ & 84 & - & - \\ \hline
Fiducial volume & 91.8\% & 76 & 1.4 &  & 77 & 0.1 & 0.05 \\ \cline{1-4}\cline{6-6}
Energy cut & 97.8\% & & & 410 &  &  &  \\ \cline{1-2}
Time cut & 99.1\% & 73 & 1.3 &  & 71 &  &  \\ \cline{1-2}\cline{5-5}
Vertex cut & 98.7\% & & & 1.1 &  &  &  \\ \cline{1-6}
Muon veto & 83\% & 60 & 1.1 & 0.9  & 1.6 &  &  \\ \hline
Combined & 73\% & 60  & \multicolumn{5}{c|}{3.8} \\ \hline
\hline
\end{tabular}
\caption{Signal and backgrounds daily rates, together with the efficiencies of the antineutrino selection criteria. From~\cite{yellow_book}.
\label{tab:sigbkg}}
\end{table}

The four neutrino mixing parameters relevant to describe the \juno{} energy spectrum are 
$\theta_{13}$, $\theta_{12}$, $\Delta m^2 _{12}$, and $\Delta m^2_{ee}$ (a definition of $\Delta m^2_{ee}$ can be found 
in the supplement material of \cite{db}). 
\juno{} won't be able to outperform Daya Bay in the determination of $\theta_{13}$ since its baseline is optimised to be
at the solar oscillation maximum, but it is expected to carry out a measurement
of the remaining three parameters with a sub-percent precision. A detailed $\chi^2$-based sensitivity analysis 
shows that, after considering shape and normalisation uncertainties due to both background and detector response, 
six years of data allow to determine $\theta_{12}$, $\Delta m^2 _{12}$, and $\Delta m^2_{ee}$ with a precision of
0.67\%, 0.59\%, 0.44\% respectively. It is worth stressing that the determination of the solar parameters does not 
depend on the energy resolution, because the solar oscillation manifests itself through broad features in energy spectrum.
As a consequence, such measurement can be performed even while the detector response is still being commissioned 
---early stage of the experiment--- and, more importantly, can also be performed  with the large- and small-\pmt{} systems
independently (the latter features a \textasciitilde10\% energy resolution), allowing a powerful validation of the
solar parameters determination.

\noindent\begin{tabular*}{\textwidth}{ p{0.45\textwidth} p{0.02\textwidth} p{0.45\textwidth}   }
\includegraphics[height=5cm]{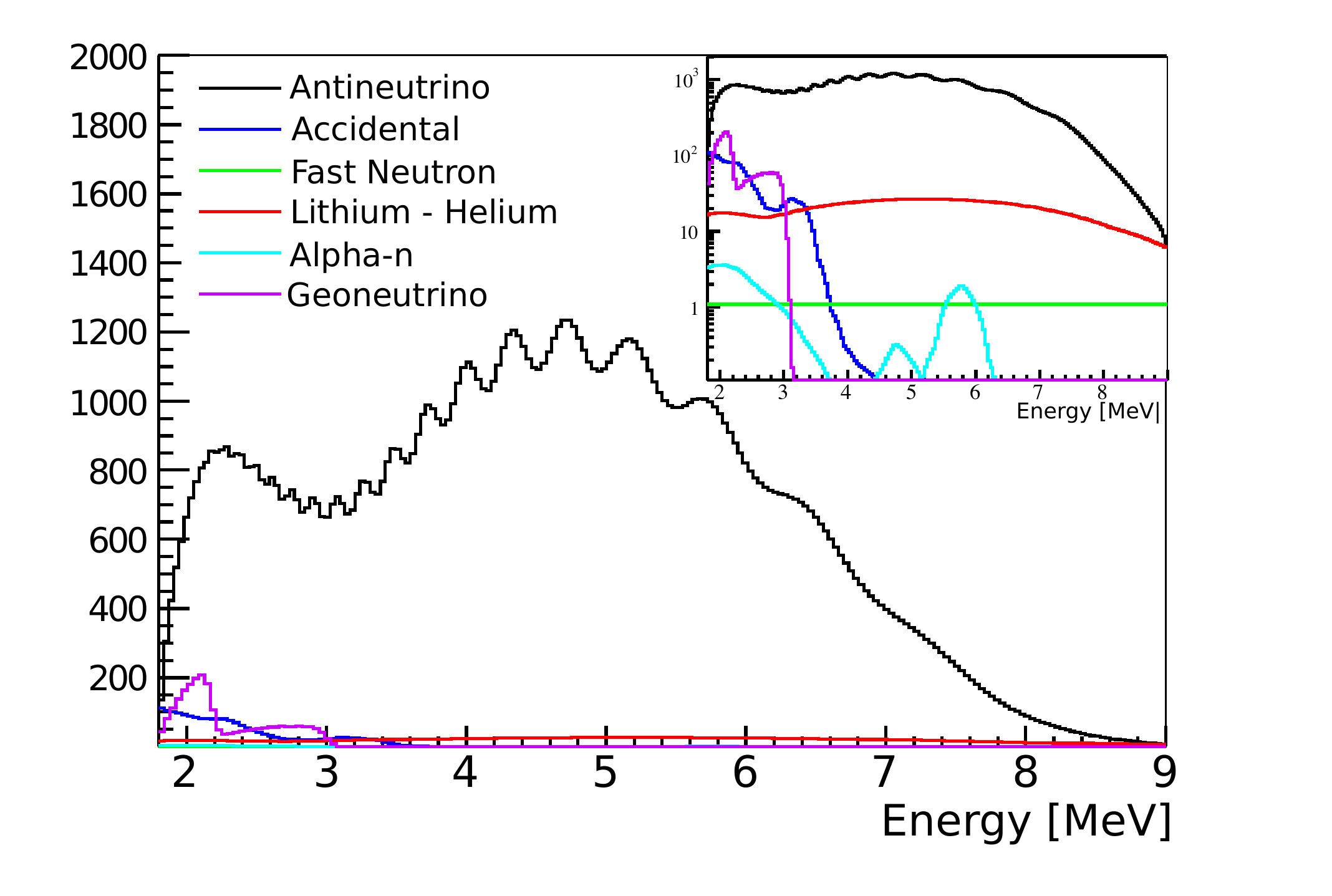}
&
&
\includegraphics[height=5cm]{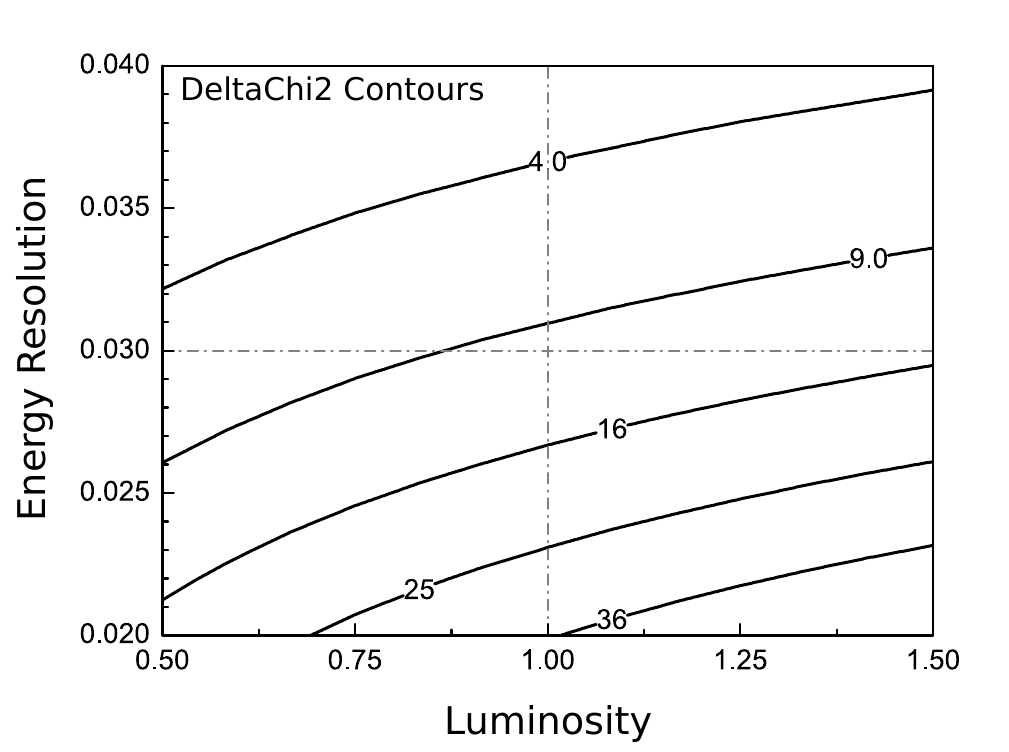} \\
{\small \textbf{Figure 1.} 
Energy spectra of the antineutrino signal and of the main background
sources: accidental coincidences, cosmogenic $^8$He and $^9$Li, 
fast neutrons, 
$^{13}$C$(\alpha, n)^{16}$O, and geoneutrinos. From~\cite{yellow_book}.
\vspace{10pt}}
& &
{ \small \textbf{Figure 2.}
\juno{}'s sensitivity to the mass ordering determination
as a function of the event statistics (luminosity) and of the
detector energy resolution. Solid lines represent 
iso-$\Delta \chi^2$ contours. Dashed lines show the 
nominal runtime of six years (including 80\%  efficiency)
and the nominal 3\% energy resolution. From~\cite{yellow_book}.
}\\
\end{tabular*}

\vspace{10pt}
\noindent \textsc{ \textbf{Supernova Burst Neutrinos}}\\
\indent A supernova SN is a stellar explosion that briefly outshines an entire galaxy, 
radiating as much energy as the Sun or any ordinary star is expected 
to emit over its entire life span. During such explosion, 99\% of 
the gravitational binding energy of the newly formed neutron 
star is emitted in the form of neutrinos.
The observation of SN neutrinos is expected to play a relevant role both in  
particle physics and astrophysics. Here we focus on the latter, where a SN signal might
help answering several fundamental questions, such as (\textsc{i}) what are the conditions
inside massive stars during their evolution? (\textsc{ii})
what mechanism triggers the SN explosion? (\textsc{iii})
are SN explosions responsible for the production of heavy chemical elements? and
(\textsc{iv}) is the compact remnant a neutron star or a black hole?
Each of these questions would deserve a dedicated section, but because of the limited space
we consider (\textsc{i}) as a case study.

The Standard Stellar Evolution Model describes temperature and density of a star
as a function of time and distance from its centre. Optical observations usually provide 
benchmark data to test it, but they have little power in constraining
the model of the star's innermost layers. Indeed, the star's high density results in optical photons 
propagating mainly via diffusion, hence losing all the information about the stellar core.
On the contrary, neutrinos interact weakly with stellar matter, and they represent a
powerful tool to probe the inner structure of the star.
For a star close to its collapse, neutrino production is dominated by thermal processes
(mainly $e^{+}$-$e^{-}$ annihilating into $\nu$-$\overline{\nu}$ pairs). That is, the neutrino
production rate, and the neutrino mean energy, both increase significantly with temperature. As a result, 
the last stages of the star's nuclear burning produce the most abundant neutrino signal 
(called pre-SN $\nu$), easier to detect and powerful in describing the stellar evolution.

Fig.~3 shows the simulated inverse beta decay (IBD) event rate in \juno{} for the nearest possible SN progenitor 
(the red supergiant Betelgeuse) whose mass is taken to be 20 solar masses (M$_{\odot}$) at a distance of 0.2~kpc. 
The sudden drop in the rate around 0.6 day before the SN explosion is ascribable to a drop 
of the core temperature, mostly due to the silicon depletion of the core itself.  
In the case of a SN explosion, \juno{}'s capability to precisely measure the position of such a dip 
%Measuring precise position of such a dip could therefore
could serve as a discriminator for different progenitor star masses. 
Moreover, the quick rise starting few hours prior to core collapse 
makes \juno{} an ultimate pre-warning system of SN explosion,
extremely valuable to the astrophysics community.

For a typical galactic SN at 10~kpc, there will be more than 5000 signal events solely from the IBD channel.
However, several other neutrino interactions contribute to the total event rate.
They differ in terms of total yield, energy spectrum and energy threshold.
Fig.~4 shows all of them together, where $(\mathrm{E_d})$
is the deposited visible energy in the detector, $(\mathrm{E^{th}})$ is the energy threshold of
each process, 
$(\nu\text{-p})$     are the neutral current interactions on protons, 
$(\nu\text{-e})$     are the elastic scatterings on electrons,
$(^{12}\text{C NC})$ are the neutral-current-mediated carbon excitations,
$(^{12}\text{N CC})$ are the charged current \nue{} interactions on $^{12}$C, and
$(^{12}\text{B CC})$ are the the same charged current interactions initiated by \anue{}.

\noindent\begin{tabular*}{\textwidth}{ p{0.45\textwidth} p{0.02\textwidth} p{0.45\textwidth}   }
\includegraphics[height=5cm]{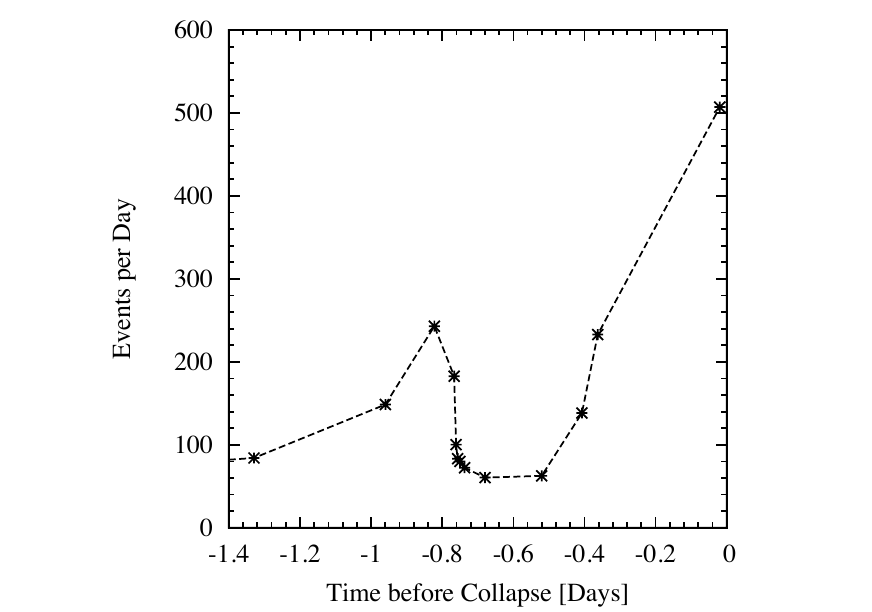}
&
&
\includegraphics[height=5cm]{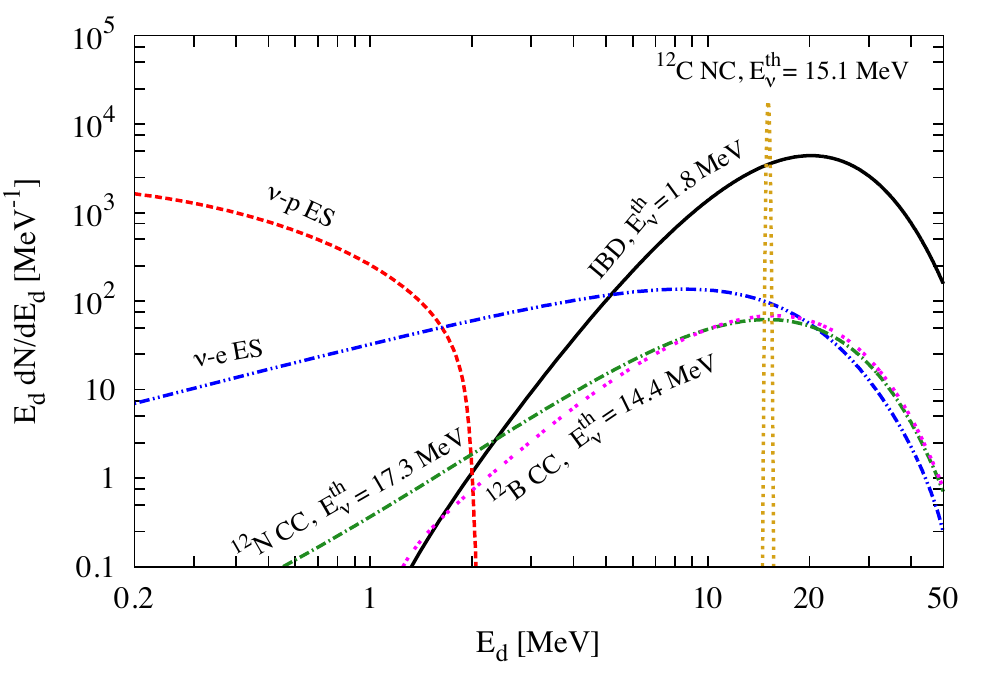} \\
{\small \textbf{Figure 3.} Neutrino event rate in \juno{} for a massive star (20 
M$_{\odot}$) distant 0.2~kpc from the Earth, 
the same as the nearest possible SN progenitor Betelgeuse. From~\cite{yellow_book}. \vspace{10pt}}
& &
{ \small \textbf{Figure 4.} Neutrino energy spectra 
%with respect to the visible energy $\mathrm{E_d}$ 
in the \juno{} detector for a SN at 10~kpc, where no neutrino flavor conversions is assumed. 
%The main reaction channels are shown together with the threshold of neutrino energies.
$\mathrm{E_d}$ is the visible energy, and $\mathrm{E^{th}}$ the threshold energy. From~\cite{yellow_book}.
}\\
\end{tabular*}

%\section{Solar Neutrinos}
\noindent \textsc{ \textbf{Solar Neutrinos}}\\
\indent The Sun is a powerful source of \nue{} with O(1~MeV) energy, produced in the thermonuclear 
fusion reactions happening in the solar core. \juno{}'s solar neutrino  programme focuses on those emitted by the 
$^7$Be and $^8$B chains.
Indeed, despite the great achievements of the last decades, there are still important aspects of solar 
neutrino physics to clarify, and some questions of great relevance for astrophysics and elementary 
particle physics waiting for definite solutions. Two of the most important ones are
(\textsc{i}) the solution of the solar metallicity problem, 
and (\textsc{ii}) the detailed analysis of the 
oscillation-probability energy dependence in the lower end 
of the $^8$B neutrino spectrum.

(\textsc{i}) The solar metallicity problem emerged when
the former agreement between Standard Solar Model (SSM) and solar data 
got compromised by the revision of the solar surface heavy element content, 
leading to a discrepancy between the SSM and helioseismology results.
The predictions of different SSM versions differ (also) by the $^8$B and $^7$Be neutrino fluxes.
\juno{}'s capability to determine these fluxes with high accuracy,
%Therefore, a possible improvement at JUNO of the accuracy in the determination of these fluxes, 
together with data (coming from other future experiments) about the CNO fluxes, 
could help solving this key issue in nuclear astrophysics.

(\textsc{ii}) According to the Mikheyev-Smirnov-Wolfenstein (MSW) effect, neutrino oscillation
parameters are different if a neutrino propagates through matter or in vacuum.
In the case of solar \nue{},
the transition between the two behaviors %the vacuum-oscillation and matter-oscillation dominated regions 
is expected to happen in the 1\textasciitilde3~MeV energy range, therefore
 solar $^8$B neutrinos ---with their  continuous energy spectrum stretching far beyond 3~MeV ---
are a privileged tool to study the MSW-modulated energy dependence.
%Indeed, according to the standard LMA-MSW solution one would expect a 
The theory predicts a smooth transition between 
the vacuum and matter related \nue{}-survival probabilities,
namely an up-turn in the spectrum. However,
none of the existing experiments so far observed a clear evidence of this effect.
% i.e. an up-turn in the spectrum.
%the low energy part of the solar $^8$B spectrum, which gave rise to a series of theoretical discussions of 
%sub-leading non-standard effects of the neutrino survival probability. 
The only exception is Super-Kamiokande, which got a mild evidence of the up-turn in its data~\cite{Renshaw:2013dzu}.
%which is slightly pointing towards the presence of the up-turn.
\juno{}'s capability to perform an independent and  high-significance 
test of the up-turn existence would be extremely important to confirm the consistency 
of the standard LMA-MSW solution, or to indicate any possible deviations from this standard paradigm.

\noindent\begin{tabular*}{\textwidth}{ p{0.45\textwidth} p{0.02\textwidth} p{0.45\textwidth}   }
   \includegraphics[height=5cm]{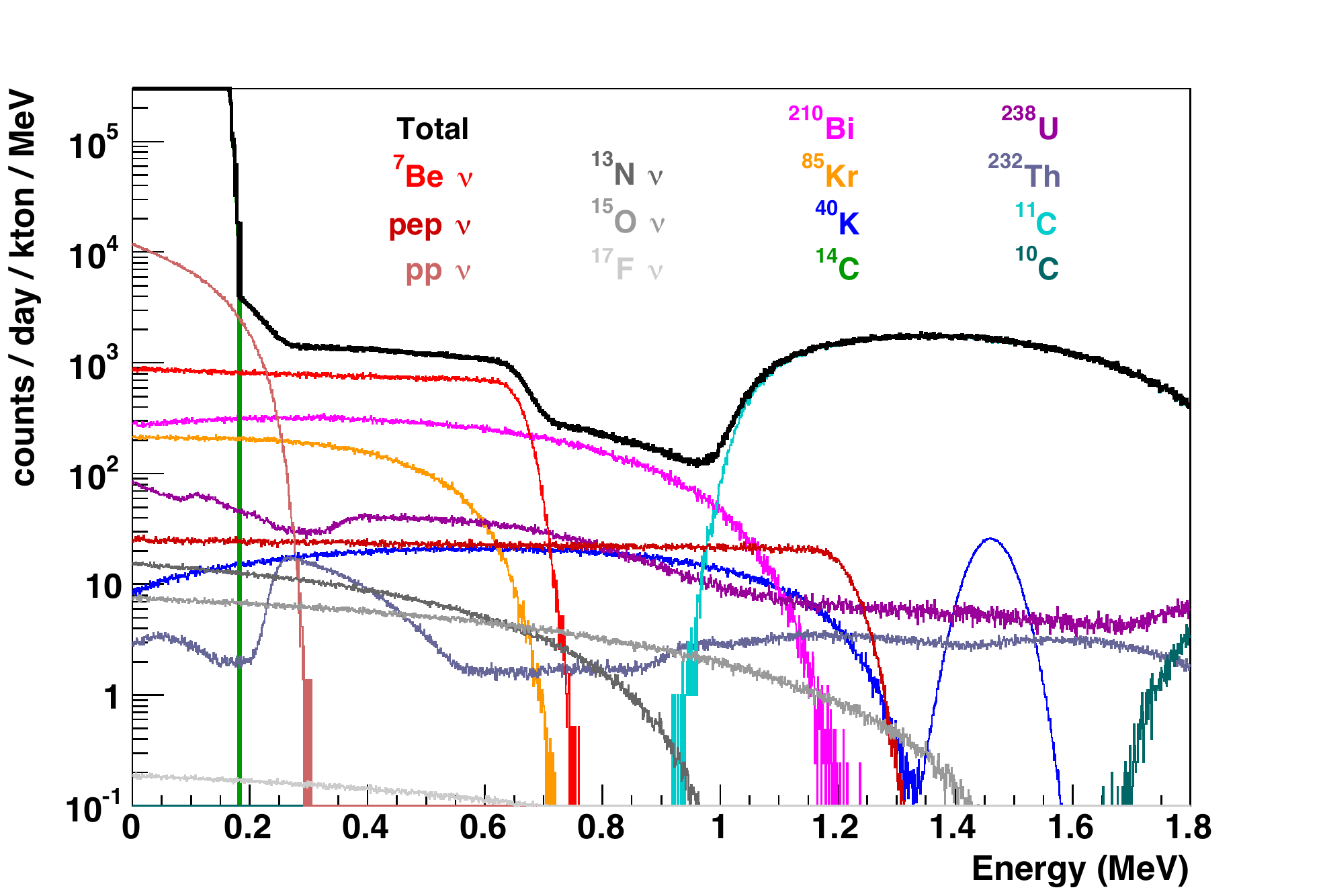}
&
&
\includegraphics[height=5cm]{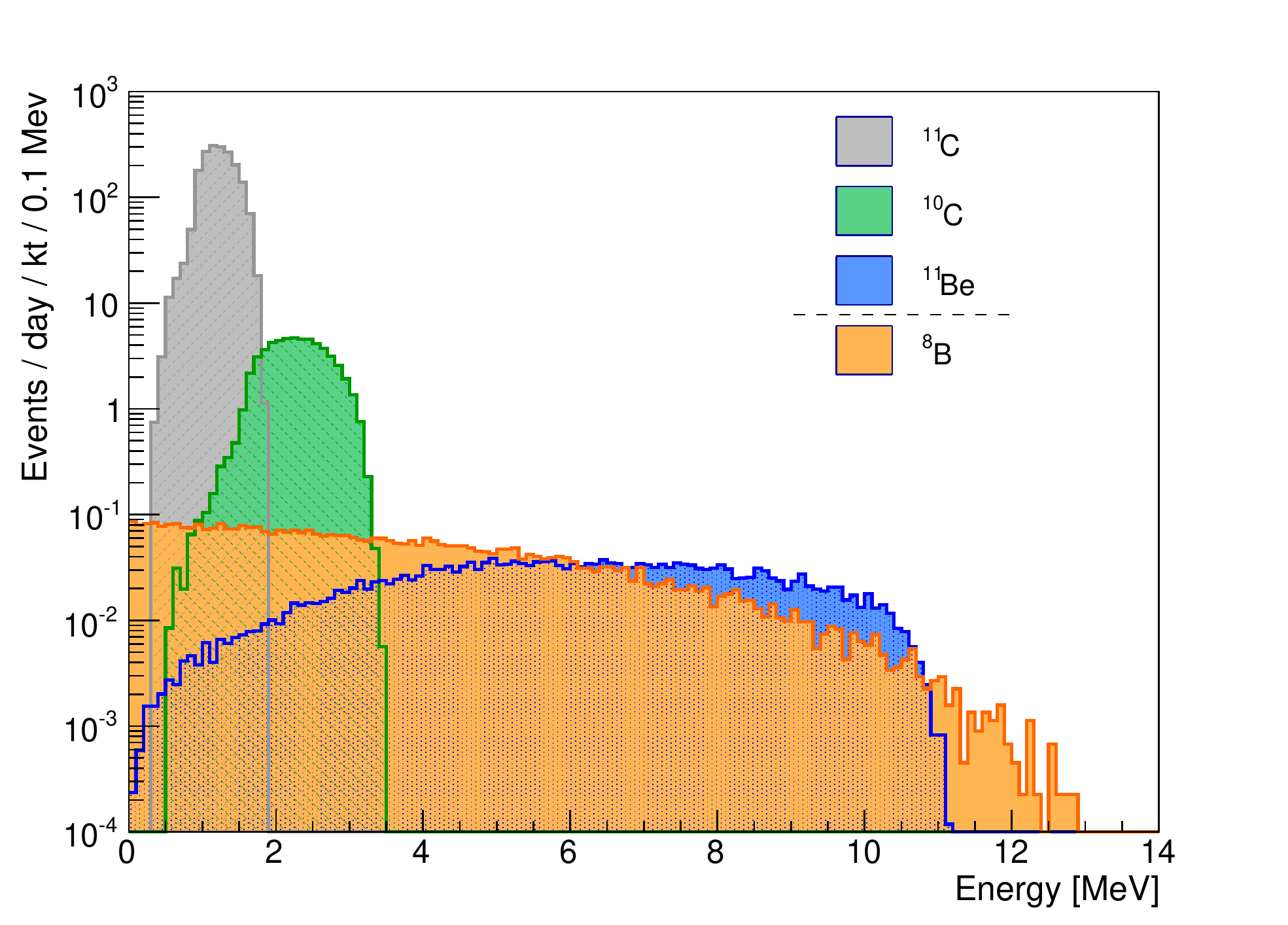}
\\
   {\small
   \textbf{Figure 5.} Energy spectra of singles from natural radioactivity (background) in the \textit{ideal}
   radiopurity scenario, together with the $^7$Be neutrino signal. From~\cite{yellow_book}. \vspace{10pt}} 
& &
  { \small \textbf{Figure 6.} 
   Energy spectra of the $^8$B neutrino signal and of the cosmogenic backgrounds. From~\cite{yellow_book}.}\\

\end{tabular*}

The challenge in detecting solar \nue{} at \juno{} is that they are detected only by elastic scattering,
which results in an experimental signature (single energy deposition) indistinguishable from most
of the background processes. The two main background sources are natural radioactivity and cosmogenic isotopes.
The first needs to be suppressed by achieving a high-level radiopurity in all the detector components. 
\textsc{Juno}'s \textit{baseline} radiopurity scheme envisages a $^{232}$Th, $^{40}$K, $^{14}$C residual contamination at
the level of $10^{-16}$, $10^{-16}$,  $10^{-17}$ g/g respectively. As a comparison, the same level of
radiopurity was achieved during KamLAND solar phase, and it would allow \juno{} to achieve a signal/background
ratio of 1/3. A more demanding radiopurity scheme (called \textit{ideal}) requires the previous contamination levels to improve
by one order of magnitude, which would correspond to Borexino phase I, and would allow a 2/1 signal/background ratio.
The energy spectra of the radioactive background processes in the case of \textit{ideal} radiopurity, together with the signal 
$^7$Be \nue{}, are shown in Fig.~5.
Among the cosmogenic isotopes, the most dangerous are the long-lived $^{11}$C ($\tau=24.4$~min) 
, $^{10}$C ($\tau = 27.8$~s), and $^{11}$Be ($\tau=19.9$~s), since they cannot be suppressed by a muon veto
without introducing large deadtime. The energy spectra of these background events are shown in 
Fig.~6. The only way to handle them is to tag them via a three-fold coincidence (muon + spallation neutron
+ isotope decay) and subtract them statistically from the total spectrum.

%\section{Geoneutrinos}
\vspace{10pt}
\noindent \textsc{ \textbf{Geoeutrinos}}\\
\indent Over the last half a century, the Earth's surface heat flow has been established to be $(46\pm3)$~TW~\cite{heat}. 
However the community is still vigorously debating what fraction of this power comes from primordial 
versus radioactive sources. This debate touches on the composition of the Earth, the question of chemical 
layering in the mantle, the nature of mantle convection, the energy needed to drive plate tectonics, 
and the power source of the geodynamo, which powers the magnetosphere that shields the Earth from the harmful cosmic ray flux.
Radioactive beta-decays of heavy elements (such as Th and U) taking place inside the Earth 
result in an upwards \anue{} flux (also called geoneutrino flux) which can be detected at \juno{} by means
of IBD reactions. A precise measurement of such flux would allow us to accurately define
the absolute abundance of Th and U in the Earth, which in turn would allow us to:
(\textsc{i}) define the building blocks, the chondritic meteorites, that formed the Earth,
(\textsc{ii}) discriminate models of parameterised mantle convection that define the thermal evolution of the Earth,
(\textsc{iii}) potentially identify and characterize deep, hidden reservoirs in the mantle, and
(\textsc{iv}) fix the radiogenic contribution to the terrestrial heat flow.
Moreover, such studies can place stringent limits on the power of any natural nuclear reactor in or
near the Earth's core.

The main experimental challenge in detecting a geoneutrino signal is
to disentangle it from the reactor \anue{} signal, which is overwhelming. Such a separation can be done
only via statistical subtraction, and it relies heavily on a precise modeling of the 
low-energy reactor \anue{} spectrum. Moreover, to interpret the geoneutrino signal in terms of mantle's
radioactivity, the contribution from the Earth's crust need to be subtracted, 
since it has been well established that
the crust surrounding the detector will play a major role in total geoneutrino budget. 
Thus, to understand the relative contributions from the crust and mantle to the total 
geoneutrino signal at \juno{}, detailed geological, geochemical, and geophysical studies 
need to be performed in the areas surrounding the detector.

\section{The JUNO Detector}
The \juno{} detector is designed to be placed 720~m underground and comprises several components. 
We refer to the Central Detector as the 35.4 m wide acrylic sphere
containing 20~kt of purified LAB scintillator (target volume). The target volume is monitored by 17000 20-inch \pmt{}s and 
34000 3-inch \pmt{}s, installed on a Stainless Steel Lattice Shell surrounding the acrylic sphere at a distance of few meters.
The overall photocathode density is the largest ever built, and accounts for an unprecedented 78\% photocoverage, yielding
1200 photoelectrons/MeV, pivotal to achieve the 3\% energy resolution at 1 MeV. The set of large \pmt{}s in unevenly split
between dynode-based Hamamatsu Photonics \pmt{}s, and  Micro Channel Plate-based \pmt{}s produced by the Chinese North Night 
Vision Technology, the latter accounting for three fourth of the total.  
The whole Central Detector is immersed
in a cylindrical water pool filled with ultra-pure water. Since the water penetrates the Lattice Shell supporting the \pmt{}s,
it acts as a buffer shielding the target volume from natural radioactivity arising from the rock and from the \pmt{} glass.
The water pool is instrumented with 2000 20-inch \pmt{}s to detect Cherenkov light, hence acting as an active muon veto.
To better track cosmic muons, a tracker consisting of three layer of plastic scintillator (inherited from the OPERA detector~\cite{opera}) 
is deployed on top of the water pool,
covering  \textasciitilde50\% of the water pool surface. The chimney connecting the acrylic sphere to the surface of the water 
pool is also instrumented in order to detect stopping muons that might generate untagged background events.

\begin{center}
   \includegraphics[height=5.7cm]{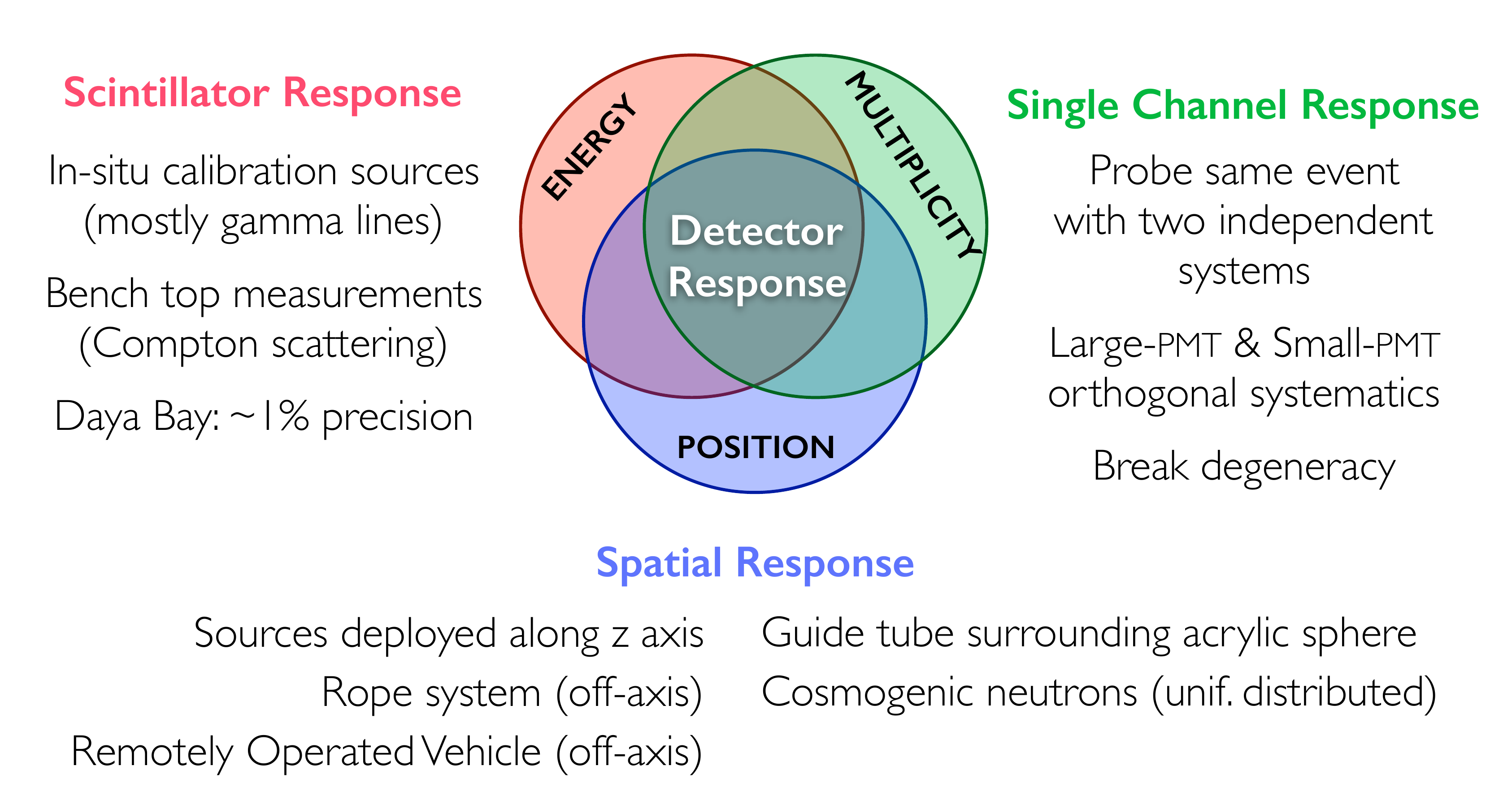} \\
   {\small
   \textbf{Figure 7.} Summary of the methods put in place to understand the detector response. }\\
\end{center}

The unprecedented light level of 1200 photoelectrons/MeV allows us to achieve a stochastic resolution term smaller than $2.9\%/\sqrt{E}$.
%This  is possibile not only through a large photocoverage, but also because a LS purification plant has been designed to 
%
%LAB purification plays a key role in achieving a LS attenuation length greater than 20~m. Great transparency is in turn
%complementary to large photocoverage to detect 1200 photoelectrons/MeV, hence minimising the resolution stochastic term.
%
%pivotal o achieve the unprecedented design resolution of 3\% at 1~MeV, the resolution stochastic term need to be 
%
However, the design resolution of 3\% at 1~MeV requires to keep non-stochastic components of the energy 
resolution below 1\%. These components mostly arise from an imperfect knowledge of the detector response, and can be 
grouped into energy-related, multiplicity-related and position-related (see Fig.~7). The latter are addressed by scanning the detector with 
four calibration systems: \textsc{(i)} one along the z axis, \textsc{(ii)} one based on ropes deployed from the chimney and able to reach most of 
the off-axis region, \textsc{(iii)} one based on a guide-tube running along the acrylic vessel to study the outermost shell of the target volume, and 
\textsc{(iv)} one based on a Remote-Operated Vehicle to check eventual blind spots. The energy-related uncertainties are addressed
by deploying radioactive sources yielding gamma lines at different energies and neutrons, as well as studying the LS response
with bench-top experiments. Finally, the multiplicity-related uncertainties are addressed by means of the small-\pmt{} system, where
the size of the \pmt{}s ensures that each phototube works in photon-counting mode ---i.e. it detects one photoelectron at most---,
effectively resulting in a \textit{digital} calorimetry estimator. This estimator is designed to be much more robust to uncertainties 
%usually induced by collecting light with 
typical of large phototubes, such as low-resilience to magnetic field, uniformity of the photocathode deposition,
knowledge of the collection efficiency profile, pileup of multiple hits, complex waveform reconstruction, and saturation
of the output signal. 
The small-\pmt{} system further provides a powerful handle to help break the degeneracy existing among the three groups, 
namely the cross-talk existing among different systematic uncertainties. As an example, an imperfect reconstruction
of those \pmt{} waveforms experiencing large pile-up might easily mimic a residual detector non-uniformity. Such degeneracy
could likely become the main limitation to a full understanding of the detector response.
The study of calibration events with both the large- and the small-\pmt{} systems, each of them providing an independent energy 
measurement
%But the small-\pmt{}, by providing an energy measurement independent from the large-\pmt{} one,
%
%The small-\pmt{} system provides for each event an energy measurement independent from the large-\pmt{} one,
%
characterised by its own hardware and reconstruction method, allows us to sample the same energy deposition
with effectively two detectors, each one experiencing a completely different systematic uncertainty budget.
The comparison and the cross-calibration of these two systems 
%The small-\pmt{}  energy estimator 
is therefore geared to become a valuable asset in understanding and minimising the non-stochastic resolution terms.

\section{Conclusions}
The Jiangmen Underground Neutrino Observatory (\juno{}) is a large and high precision liquid scintillator detector
under construction in the south of China. With its 20~kt target mass, it aims to achieve an unprecedented 3\% 
energy resolution at 1~MeV. To this end, a 1200 photoelectrons/MeV light level is required, which drives the effort
to reach a LS attenuation length larger than 20~m, and a photocoverage larger than 75\%. Such stringent 
requirements are pivotal to determine the neutrino mass ordering through the analysis of \anue{} produced
by two powerful nuclear power plants (36~GW$_{\mathrm{th}}$ nominal power) at a baseline of 53~km.
The investigation of  reactor \anue{} is also aimed to determine the neutrino mixing parameters $\theta_{12}$, $\Delta m^2_{12}$, and
$\Delta m ^2_{ee}$ with a precision better than 1\%. \juno{} will further be able to detect neutrinos coming from
a supernova burst, and geoneutrinos originating from the inner layers of our planet, hence extending \juno{}'s
Physics Programme far beyond the reactor neutrino physics. Depending on the scintillator
radiopurity, neutrinos originating from the Sun could also be detected, shedding light on the solar 
metallicity problem and on the Mikheyev-Smirnov-Wolfenstein turn on curve. \juno{}'s data taking is foreseen to begin 
in 2020.


\begin{thebibliography}{99}

\bibitem{CDR}
\newblock F.~An {\it et al.} [JUNO Collaboration],
\newblock \emph{JUNO Conceptual Design Report}.
\newblock [\textit{arXiv:}~\href{http://arxiv.org/abs/1508.07166}{1508.07166}]

\bibitem{pmns}
Z.Maki {\it et al.}, Prog. Theor. Phys. {\bf 28} (1962) 870, 
B.Pontecorvo, Zh. Eksp. Theor. Fis., {\bf 53} (1967) 1717, 
V.Gribov and B.Pontecorvo, Phys. Lett. B {\bf 28} (1969) 493.

\bibitem{yellow_book}
  F.~An {\it et al.} [JUNO Collaboration],
  %``Neutrino Physics with JUNO,''
  J.\ Phys.\ G {\bf 43} (2016) no.3,  030401
  %doi:10.1088/0954-3899/43/3/030401
  [arXiv:1507.05613].
  %%CITATION = doi:10.1088/0954-3899/43/3/030401;%%
  %57 citations counted in INSPIRE as of 18 Aug 2016

\bibitem{kamland}
S. Abe \textit{et al.} (KamLAND),
%"Precision Measurement of Neutrino Oscillation Parameters with KamLAND",
Phys. Rev. Lett. \textbf{100} (2008) 221803,
[arXiv:0801.4589].

\bibitem{db}
F. P. An \textit{et al.} (Daya Bay),
%"A new measurement of antineutrino oscillation with the full detector configuration at Daya Bay",
Phys. Rev. Lett. \textbf{115} (2015) 111802,
[arXiv:1505.03456].


\bibitem{dc}
Y. Abe \textit{et al.} (Double Chooz),
%"Measurement of $\theta_{13}$ in Double Chooz using neutron captures on hydrogen with novel background rejection techniques",
JHEP \textbf{01} (2016) 163,
[arXiv:1510.08937].

\bibitem{reno}
J.H. Choi \textit{et al.} (RENO),
%"Observation of Energy and Baseline Dependent Reactor Antineutrino Disappearance in the RENO Experiment",
Phys.Rev.Lett. \textbf{116} (2016) 211801,
[arXiv:1511.05849].

\bibitem{Zhan:2009rs}
  L.~Zhan, Y.~Wang, J.~Cao and L.~Wen,
  %``Experimental Requirements to Determine the Neutrino Mass Hierarchy Using Reactor Neutrinos,''
  Phys.\ Rev.\ D {\bf 79} (2009) 073007
  %doi:10.1103/PhysRevD.79.073007
  [arXiv:0901.2976].
  %%CITATION = doi:10.1103/PhysRevD.79.073007;%%
  %76 citations counted in INSPIRE as of 18 Aug 2016

\bibitem{Renshaw:2013dzu}
\newblock A.~Renshaw {\it et al.} [Super-Kamiokande Collaboration],
%\newblock \emph{First Indication of Terrestrial Matter Effects on Solar Neutrino Oscillation}.
\newblock Phys.\ Rev.\ Lett.\  {\bf 112} (2014) 9,  091805.

\bibitem{minos}
P. Adamson \textit{et al.} (MINOS),
%"Combined analysis of $\nu_{\mu}$ disappearance and $\nu_{\mu} \rightarrow \nu_{e}$ appearance in MINOS using accelerator and atmospheric neutrinos",
Phys.Rev.Lett. \textbf{112} (2014) 191801,
[arXiv:1403.0867].

\bibitem{heat}
J. H. Davies and D. R. Davies, Solid Earth \textbf{1} (2010) 5.

\bibitem{opera}
R. Acquafredda \textit{et al.} (OPERA), JINST \textbf{4}
(2009) P04018.

\end{thebibliography}
\end{document}